\documentclass[prl,twocolumn,superscriptaddress,%
preprintnumbers,amsmath,amssymb]{revtex4}
\usepackage{amsmath}
\usepackage{graphicx}
\usepackage{dcolumn}
\DeclareGraphicsExtensions{.png .jpg .pdf}

\begin{document}

\title{Landau levels of single layer and bilayer phosphorene}

\author{J. M. Pereira Jr.}\email{pereira@fisica.ufc.br}
\affiliation{Departamento de F\'isica, Universidade
Federal do Cear\'a, Caixa Postal 6030, Campus do Pici, 60455-900
Fortaleza, Cear\'a, Brazil}
\affiliation{Institute for Molecules and Materials, Radboud University Nijmegen,
Heijndaalseweg 135, 6525 AJ, Nijmegen, The Netherlands}
\author{M. I. Katsnelson}%\email{pereira@fisica.ufc.br}
\affiliation{Institute for Molecules and Materials, Radboud University Nijmegen,
Heijndaalseweg 135, 6525 AJ, Nijmegen, The Netherlands}

\date{ \today }

\begin{abstract}
In this work we introduce a low-energy Hamiltonian for single layer and bilayer black phosphorus that describes the electronic states at the vicinity of the gamma point. The model is based on a recently proposed tight-binding description for electron and hole bands close to the Fermi level. We calculate expressions for the Landau level spectrum as function of magnetic field and in the case of bilayer black phosphorus we investigate the effect of an external bias on the electronic band gap. The results showcase the highly anisotropic character of black phosphorus and in particular for bilayer BP, the presence of bias allows for a field-induced semiconductor-metal transition.
\end{abstract}

\maketitle

In the last ten years the properties of crystals consisting of one or few atomic layers has been the focus of intense research. Such interest arose mainly due to the production of graphene in 2004, which has been shown to display remarkable electronic, optical and mechanical properties \cite{Misha1}. Since then, there has been a growing interest in the production of other low-dimensional crystals. The investigation of analogs of graphene has resulted in the discovery of several single layer crystals of different elements, such as Silicon (silicene) \cite{silicene}, Germanium (germanene) \cite{germanene}, as well as a class of materials known as transition metal dichalcogenides \cite{Kis}. Some of these materials may soon find use in electronic devices, mainly due to the fact that in contrast with graphene, they present a band gap in their electronic spectrum, albeit with a lower carrier mobility. Among the most promising of these 2D materials is an allotrope of Phosphorus, known as black phosphorus (BP) \cite{bp1,bp2,bp3,bp4,bp5,bp6,bp7}, which is that element's most stable crystal at room temperature and pressure. In bulk, BP is a narrow gap semiconductor with a orthorhombic structure that consists of atoms covalently bound into layers coupled by van der Waals interactions. Similarly to graphene, BP can be mechanically exfoliated to obtain samples with few or single layers, with the latter being known as phosphorene. The resulting material has a band gap that depends on the number of layers, varying from $0.6$ eV for five layers to $1.5$ eV for a single layer, with carrier mobility in the range of $\approx 1000$ cm$^2$ V$^{-1}$s$^{-1}$. 

The importance of a thorough understanding of the band structure and charge carrier dynamics in BP has led to a series of recent studies that obtained the electronic dispersion using approaches such as first principles calculations, ${\mathbf k} \cdot {\mathbf p}$ methods, as well as tight-binding models \cite{Rudenko}. These calculations have shown evidence of a large anisotropy on the effective mass, as well as given estimates of the energy gap for single and multilayer BP. Calculations have shown the possibility of a topologial phase transition in few-layer BP, in which an external bias induces a band inversion \cite{Fazzio}. This would allow the development of devices in which the topological character of the material can be externally controlled.

In this work, we consider the charge carrier dynamics in single layer and bilayer phosphorene by means of a continuum model obtained as the long wavelength limit of a recently proposed tight-binding model \cite{Rudenko}. In addition to the anisotropy of the spectrum, another striking feature of the electronic bands obtained from this model is the hybrid nature of the electron and hole states close to the band edge in phosphorene, which display both a Schr\"odinger-like and Dirac-like character, which in turn is dependent on the direction of propagation.
For the case of bilayer BP, we also consider the effect of an external bias on the spectrum. We obtain results that show a bias-induced gap closure, which leads to the presence of zero-energy Landau levels.

The paper is organized as follows: in section II we present the model Hamiltonian for single BP layers and analytical expressions for its Landau level spectrum. Section III extends that model for the case of the bilayer. Finally, in section IV we present a discussion of the results and conclusions.

\section{Single layer phosphorene}

The structure of each layer of BP has phosphorus atoms covalently coupled to three nearest neighbors. The resulting lattice resembles the honeycomb structure of graphene, however in phosphorene the sp$^3$ hybridization of the $3$s and $3$p atomic orbitals creates ridges that result in a puckered surface (Fig. 1). Using the tight-binding model proposed in Ref \cite{Rudenko}, we can write the Hamiltonian for single layer black phosphorus as
\begin{equation}
{\mathcal H}_k=
\begin{pmatrix}
  u_A & t_{AB}(k)& t_{AD}(k) & t_{AC}(k) \\
  t_{AB}(k)^* & u_B & t_{AC}(k)^* & t_{AD}(k) \\
  t_{AD}(k) & t_{AC}(k) & u_D & t_{AB}(k)\\
  t_{AC}(k)^* & t_{AD}(k) & t_{AB}(k)^* & u_C
\end{pmatrix}\quad , \quad
\end{equation}
with eigenvectors given by $[\phi_A \phi_B \phi_D \phi_C]^T$
and where $u_{A,B,C,D}$ represent the on-site energies - which we henceforth assume as equal to $U$, with the $A - C$ subscripts denoting the four sublattice labels shown in Fig. 1. The interaction terms are given in the appendix.
By taking into account the symmetries of the phosphorene lattice, one can write a reduced two-band Hamiltonian for single layer black phosphorus at the vicinity of the Fermi level as
\begin{equation}
{\mathcal H}_k=
\begin{pmatrix}
  U + t_{AD}(k) & t_{AB}(k) + t_{AC}(k)  \\
  (t_{AB}(k)+ t_{AC}(k))^* & U + t_{AD}(k)
\end{pmatrix}\quad , \quad
\end{equation}
which acts on the spinors
\begin{equation}
{\Psi}=\frac{1}{2}
\begin{pmatrix}
  \phi_A + \phi_D \\
  \phi_B + \phi_C
\end{pmatrix}\quad , \quad
\end{equation}
From the Hamiltonian Eq.(2) one can obtain the energies for the bottom of the conduction band and the top of the valence band as
$
E_c = 2t_1+t_2+2t_3+t_5 + 4t_4,
$
and
$
E_v = -(2t_1+t_2+2t_3+t_5)+4t_4.
$
That leads to a gap of $\Delta \approx 1.52$ eV. 

\begin{figure}
\includegraphics[width=8cm]{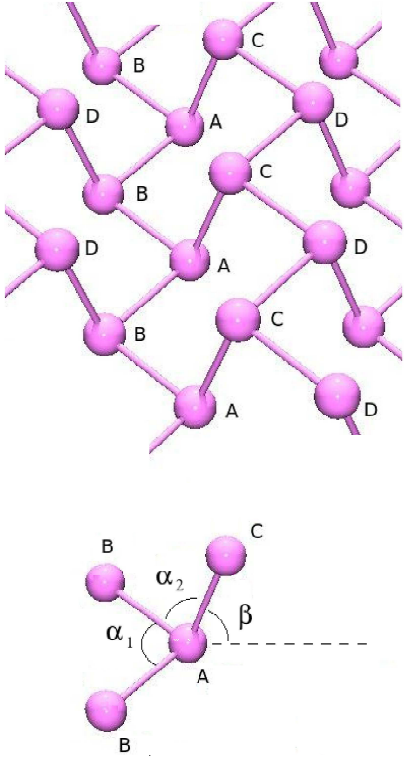}
\caption{\label{fig1} (Color online) Nearest neighbors in the phosphorene lattice. }
\end{figure}

By diagonalizing the Hamiltonian Eq. (2) one can obtain the following dispersions:
\begin{eqnarray}
E(k_x,k_y) &=& U +4t_4\cos{(k_xd_1)}\cos{(k_yd_2)}\cr
&&\cr
&&\pm \Big\{ 4[t_1^2+t_3^2+2t_1t_3\cos{(2k_yd_2)}]\cos^2{(k_xd_1)}\cr
&&\cr
&&+[t_2^2+t_5^2+2t_2t_5\cos{(2k_yd_2)}]\cr
&&\cr
&&+4t_3[t_2\cos{(k_yd_2)}+t_5\cos{(3k_yd_2)}]\cos{(k_xd_1)}\cr
&&\cr
&&+4t_1[t_2+t_5]\cos{(k_xd_1)}\cos{(k_yd_2)}\Big\}^{1/2},
\end{eqnarray}
where $d_1=a_1\sin{\alpha_1/2}$ and $d_2 =a_1\cos{\alpha_1/2}+a_2\cos{\beta}$,
with the positive (negative) sign corresponding to the conductance (valence) band. Figure 2 shows a plot of Eq. (4) centered at the gamma point (black lines), where the strong anisotropy of the spectrum is evident.

A simple calculation shows that the eigenstates of the Hamiltonian Eq.(1) can be found as
\begin{equation}
{ \Psi}_1=\frac{1}{\sqrt{2}}
\begin{pmatrix}
  1  \\
  {}\\
 \lambda e^{i\theta_k} 
\end{pmatrix}\quad ,
\end{equation}
where $\lambda = \pm 1$, with the same sign convention as Eq. (4) and
\begin{equation}
\theta_k =\tan^{-1}(C/D)
\end{equation}
where
\begin{eqnarray}
C&=&-2t_1\cos{(k_xd_1)}\sin{(k_ya_1\cos{\alpha_1/2})}+t_2\sin{(k_ya_2\cos{\beta})}\cr
&&+2t_3\cos{(k_xd_1)}
\sin{[k_y(a_1\cos(\alpha_1/2)+2a_2\cos{\beta})]}\cr
&&-t_5\sin{[k_y(2a_1\cos(\alpha_1/2)+a_2\cos{\beta})]}
\end{eqnarray}
and
\begin{eqnarray}
D&=&2t_1\cos{(k_xd_1)}\cos{(k_ya_1\cos{\alpha_1/2})}+t_2\cos{(k_ya_2\cos{\beta})}\cr
&&+2t_3\cos{(k_xd_1)}
\cos{[k_y(a_1\cos(\alpha_1/2)+2a_2\cos{\beta})]}\cr
&&+t_5\cos{[k_y(2a_1\cos(\alpha_1/2)+a_2\cos{\beta})]}.
\end{eqnarray}

Although these results show some similarity to the results for graphene it can be seen that for phosphorene the phase angle does not correspond to the polar angle of the momentum vector.

\begin{figure}
\includegraphics[width=8cm]{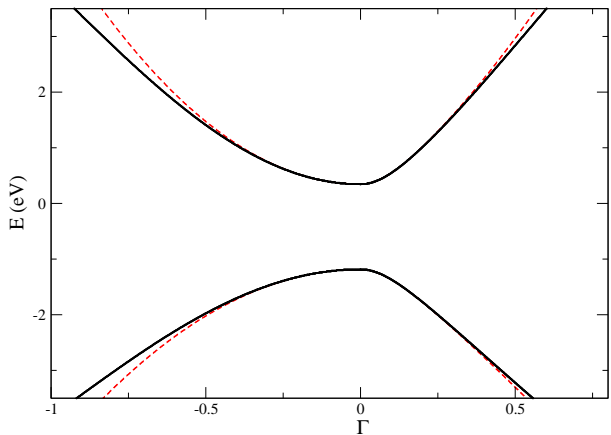}
\caption{\label{fig2} (Color online) Low-energy dispersion of phosphorene from tight-binding model (solid black lines) and continuum approximation (dashed red lines). 
}
\end{figure}

{\it Continuum approximation:} By expanding the structure factors around $k = 0$ (Gamma point) and retaining the terms up to second-order in $k$, one can write a long-wavelenght approximation for the Hamiltonian Eq.(2) as
\begin{equation}
{\mathcal H}_k=
\begin{pmatrix}
  u_0 + \eta_x k_x^2+\eta_y k_y^2 & \delta + \gamma_x k_x^2+\gamma_y k_y^2 + i\chi k_y  \\
 \delta + \gamma_x k_x^2+\gamma_y k_y^2 - i\chi k_y & u_0 + \eta_x k_x^2+\eta_y k_y^2
\end{pmatrix}\quad , \quad
\end{equation}
where 
\begin{eqnarray}
\eta_x &=& -2t_4[a_1\sin(\alpha_1 /2)]^2,\cr
\eta_y &=& -2t_4[a_1\cos(\alpha_1 /2)+a_2\cos \beta]^2,\cr
\gamma_x &=& -(t_1+t_3)[a_1 \sin(\alpha_1 /2)]^2,\cr
\gamma_y &=& -t_1[a_1\cos(\alpha_1 /2)]^2-t_3[a_1\cos(\alpha_1 /2)+2a_2\cos \beta]^2\cr
&&-t_2(a_2 \cos \beta)^2/2-t_5[2a_1\cos(\alpha_1 /2)+a_2\cos \beta]^2/2,\cr
\delta &=& t_2 + t_5 + 2(t_1+t_3),\cr
u_0&=&4t_4,\cr
\chi &=& t_2 a_2 \cos\beta + 2t_3[a_1\cos(\alpha_1 /2)+2a_2\cos\beta]-\cr  
&&t_5(2a_1\cos[\alpha_1 /2)+a_2\cos\beta]-2t_1a_1\cos(\alpha_1 /2).
\end{eqnarray}

By substituting the hopping parameters in the above expressions we obtain the following values:
 $u_0=-0.42$ eV,  
 $\eta_x=0.58$ eV$\cdot$ \AA$^2$, 
 $\eta_y=1.01$ eV$\cdot$ \AA$^2$,  
 $\delta=0.76$ eV,  
 $\chi=5.25$ eV$\cdot$ \AA, 
 $\gamma_x=3.93$ eV$\cdot$ \AA$^2$, and
 $\gamma_y=3.83$ eV$\cdot$ \AA$^2$.
 
The eigenvectors are $[\phi_1  \quad \phi_2]^T$, 
with the $\phi_{1,2}$ spinor components now corresponding to envelope functions associated with linear combinations of the amplitudes for each sublattice site. The form of Hamiltonian Eq. (9) is similar to the one presented in Ref. [7], which was obtained within the ${\mathbf k}\cdot {\bf p}$ approximation with parameters chosen in order to fit the band structure obtained from first principle calculations. In the present case, however, the parameters include the contribution from different hopping terms between neighboring lattice sites, as well as the lattice geometry, and thus can be understood as presenting a direct link between the microscopic tight-binding description and the continuum approximation. Moreover, within this model additional momentum-dependent terms can be added to better approximate the spectrum at higher energies by including higher-order $k$ terms in the structure factor expansion.
Dispersion relations for electrons and holes are then given by
\begin{equation}
E = u_0 + \eta_x k_x^2 + \eta_y k_y^2 \pm \sqrt{(\delta + \gamma_x k_x^2
+\gamma_y k_y^2)^2+\chi^2 k_y^2},
\end{equation}
where the plus (minus) sign yields the conduction (valence) band. As shown in Fig. 2, there is good agreement between the continuum and the tight-binding results for energies in the range $-2.0$ to $1.5$ eV. It can be seen, from the spectrum of Eq. (11) that, although BP has an anisotropic dispersion, it does not correspond exactly to the spectrum of a simple anisotropic system with an elliptical dispersion, due to the additional term proportional to $\chi^2$ in the radical. However, as shown below, for low energies a simple anisotropy on the effective mass can be recovered as an approximation.

{\it Effective masses:} From the spectrum Eq. (11) one can estimate the effective masses of electrons and holes in BP. Taking into account the anisotropy of the system, one can readily find, for the $x$ direction:
\begin{equation}
m^e_x = \frac{\hbar^2}{2(\eta_x + \gamma_x)}, \qquad m^h_x = \frac{\hbar^2}{2(\gamma_x - \eta_x)}.
\end{equation} 
For $m_y$ one finds, for small values of $k_y$,
\begin{equation}
m_y^{e,h}=  \frac{\hbar^2}{2(\eta_y \pm\gamma_y \pm \chi^2/2\delta)},
\end{equation}
where the plus (minus) sign corresponds to electrons (holes).
The resulting effective masses are $m^e_x = 0.846$ $m_0$ and $m^h_x = 1.14$ $m_0$, $m^e_y=0.166$ $m_0$ and $m^h_y=0.182$  $m_0$, with m$_0$ being the mass of a free electron. In comparison, the values of effective masses quoted in Ref. [8] are 
 $m^e_x = 0.7$ $m_0$ and $m^h_x = 1.0$ $m_0$, and  $m^e_y = m^h_y = 0.15$ $m_0$ (in that case, the choices of $x$ and $y$ labels were the opposite of ours).  
One then can use these masses to obtain an approximation for the spectrum Eq. (11) as (for electrons):
\begin{equation}
E = (u_0+\delta) + \frac{\hbar^2}{2m^e_x} k_x^2 + \frac{\hbar^2}{2m^e_y} k_y^2,
\end{equation}
and a corresponding expression for holes.

{\it Eigenstates:} The continuum approximation Hamiltonian Eq.(9) can be rewritten in a more compact form as
\begin{equation}
{\mathcal H}=
\begin{pmatrix}
  \epsilon_1 & \epsilon_2 e^{i\theta_k}  \\
\epsilon_2 e^{-i\theta_k} & \epsilon_1
\end{pmatrix}\quad , \quad
\end{equation}
where 
\begin{equation}
\epsilon_1=\frac{f_++f_-}{2}, \qquad \epsilon_2
=\sqrt{\Big(\frac{f_+-f_-}{2}\Big)^2 + (\chi k_y)^2},
\end{equation}
and 
\begin{equation}
\theta_k = \tan^{-1}[2\chi k_y/(f_+-f_-)],
\end{equation}
where we defined
\begin{equation}
f_{\pm} = (u_0\pm\delta) + (\eta_x \pm \gamma_x ) k_x^2+(\eta_y \pm \gamma_y) k_y^2,
\end{equation}
where, for $k_y =0$, the $f_+$ and $f_-$ expressions yield the dispersions for the conduction and valence bands, respectively.
Thus, using this polar notation, one can readily obtain the eigenstates as
\begin{equation}
{ \Psi}_{\lambda}=\frac{1}{\sqrt{2}}
\begin{pmatrix}
  1  \\
  {}\\
 \lambda e^{i\theta_k} 
\end{pmatrix}\quad ,
\end{equation}
where $\lambda = \pm 1$, with the positive (negative) signs correspond to electrons (holes). These expressions are formally similar to the states of Eq. (5), which are valid for the whole Brillouin Zone and, as before, 
the angle $\theta_k$ does not correspond necessarily to the polar angle associated with the momentum vector.
In fact, since the denominator in Eq. (18) depends only on even powers of the momentum components, the polar angle will assume values in the range $-\theta_c < \theta_k < \theta_c$, where $\theta_c < \pi/2$ is an energy-dependent critical value corresponding to $k_x=0$. From the form of Eq. (18) it is seen that as the energy increases $\theta_c$ approaches a maximum value and then decays to zero. One consequence of that behavior is the fact that, although a pseudospin may be defined for charge carriers in phosphorene for a certain energy range, the Berry phase is nevertherless zero, due to the vanishing of the winding number around the $\Gamma$ point.  

{\it Landau levels:}
In order to calculate the Landau levels for phosphorene, let us consider the Hamiltonian Eq. (9) with a magnetic field perpendicular to the plane of the layer, and use the gauge ${\mathbf A} = (-By,0,0)$ and the substitution ${\mathbf k}\rightarrow -i\nabla$. Since the Hamiltonian does not depend on $x$, we can assume $\phi_{1,2}(x,y) = \phi_{1,2}(y)e^{ik_x x}$, with $\phi_{1}=(\phi_{A}+\phi_{D})/2$ and $\phi_{2}=(\phi_{B}+\phi_{C})/2$. Thus we obtain the following pair of coupled differential equations
\begin{eqnarray}
&&[u_0 + \eta_x (k_x + \beta y)^2 - \eta_y\frac{d^2}{d y^2}]\phi_1\cr
{}\cr
&&+[\delta + \gamma_x(k_x + \beta y)^2 - \gamma_y 
\frac{d^2}{d y^2} + 
\chi \frac{d}{d y}]\phi_2 = E \phi_1\cr
{}\cr
&&[u_0 + \eta_x (k_x + \beta y)^2 - \eta_y\frac{d^2}{d y^2}]\phi_2\cr
{}\cr
&&+[\delta + \gamma_x(k_x + \beta y)^2 - \gamma_y\frac{d^2}{d y^2} - \chi \frac{d}{d y}]\phi_1 = E \phi_2,
\end{eqnarray}
where $\beta = eB/\hbar = {\ell}_B^{-2}$, with $\ell_B$ being the magnetic length.
Let us now set $k_x=0$ without loss of generality and rewrite the Hamiltonian in terms of ladder operators, acting on the spinor components $\phi_{\pm}=(\phi_1\pm \phi_2)/\sqrt{2}$,
\begin{equation}
\alpha = \sqrt{\frac{\beta}{2}}\Big( y + \frac{1}{\beta}\frac{d}{dy}\Big), \qquad 
\alpha^{\dagger} = \sqrt{\frac{\beta}{2}}\Big( y - \frac{1}{\beta}\frac{d}{dy}\Big).
\end{equation}
Thus, we can readily obtain a Hamiltonian in terms of the $\alpha$ operators as
\begin{equation}
{\mathcal H}=\Big(\frac{{\mathcal E}_+ +{\mathcal E}_-}{2}\Big){\mathbf 1}+\Big(\frac{{\mathcal E}_+ -{\mathcal E}_-}{2}\Big)\sigma_z - \chi\sqrt{\beta/2}(\alpha - \alpha^{\dagger})\sigma_x,
\end{equation}
where ${\mathbf 1}$ is the unit matrix, $\sigma_x$ and $\sigma_z$ are Pauli matrices and
\begin{equation}
{\mathcal E}_+=u_0+\delta + 2\eta_+\beta(\alpha^{\dagger}\alpha+1/2)+\Delta_+\beta(\alpha^{\dagger}\alpha^{\dagger}+\alpha \alpha),
\end{equation}
and
\begin{equation}
{\mathcal E}_-=u_0-\delta + 2\eta_-\beta(\alpha^{\dagger}\alpha+1/2)+\Delta_-\beta(\alpha^{\dagger}\alpha^{\dagger}+\alpha \alpha),
\end{equation}
where we defined $\eta_\pm = \eta \pm \gamma$ and $\Delta_{\pm} = \Delta \eta \pm \Delta \gamma$, with $\eta = (\eta_x+\eta_y)/2$, $\gamma= (\gamma_x+\gamma_y)/2$, $\Delta \eta = (\eta_x - \eta_y)/2$ and $\Delta \gamma = (\gamma_x - \gamma_y)/2$. 
A plot of the Landau levels as function of magnetic field is shown (black dots) in Fig. 3, for electrons.% and Fig. 8 (holes).
\begin{figure}
\includegraphics[width=7cm]{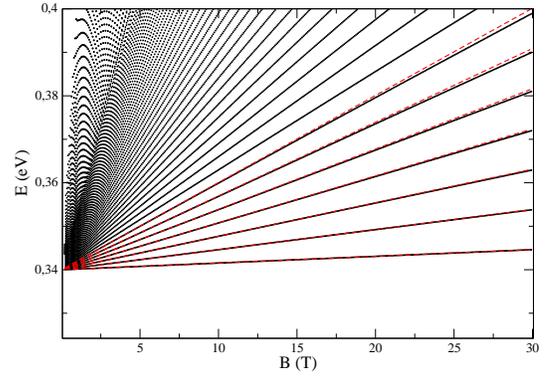}
\caption{\label{fig3} (Color online) Landau levels as function of magnetic field. The linear approximation is shown as the red dashed curves. }
\end{figure}

Although the actual spectrum deviates from the linear dependence on magnetic field for large fields, for $B < 30$ T the spectrum still shows an approximately linear dependence. In this regime, one can obtain an expression for the Landau levels by means of the following ansatz:
\begin{equation}
\phi_- = \frac{\chi}{2\delta}\sqrt{\frac{\beta}{2}}(\alpha - \alpha^\dagger)\phi_+.
\end{equation}
This ansatz can be justified by the fact that its introduction leads to an approximate Hamiltonian in which an additional term proportional to $\chi$ is added to the y-mass term (see, e.g. Eq. (13)).
Thus, using the above ansatz allows us to obtain a block diagonal Hamiltonian where the block corresponding to the electron branches is
\begin{eqnarray}
{\mathcal H}_e &=& u_0 + \delta + 2\eta_+\beta(\alpha^{\dagger}\alpha+1/2)+\Delta_+\beta(\alpha^{\dagger}\alpha^{\dagger}
+\alpha \alpha)\cr
{}\cr
&&-\frac{\chi^2}{4\delta}\beta(\alpha^{\dagger}\alpha^{\dagger}+\alpha \alpha-\alpha\alpha^{\dagger}-\alpha^{\dagger}\alpha).
\end{eqnarray}
We now define
\begin{equation}
\mu_1 = \eta_+ +\frac{\chi^2}{4\delta}, \qquad \mu_2 = \Delta_+-\frac{\chi^2}{4\delta},
\end{equation}
that allows us to rewrite the Hamiltonian Eq. (26) as
\begin{equation}
{\mathcal H}_e = u_0+\delta + 2\mu_1\beta(\alpha^{\dagger}\alpha+1/2)+\mu_2\beta(\alpha^{\dagger}\alpha^{\dagger}+\alpha \alpha).
\end{equation}
Next, one can perform a Bogoliubov transformation
\begin{equation}
c = w \alpha + v\alpha^{\dagger}, \qquad c^{\dagger} = w\alpha^{\dagger} + v\alpha,
\end{equation}
with $w^2 - v^2 = 1$, for which to $w = \cosh{\nu}$, $v = \sinh{\nu}$, $\tanh{2\nu}= \mu_2/\mu_1$. That gives us
\begin{equation}
w = \frac{1}{\sqrt{2}}\Big[\frac{\mu_1}{\sqrt{\mu_1^2-\mu_2^2}}+1 \Big]^{1/2}, \qquad
v = \frac{1}{\sqrt{2}}\Big[\frac{\mu_1}{\sqrt{\mu_1^2-\mu_2^2}}-1 \Big]^{1/2}.
\end{equation}
Finally, one can readily obtain the transformed Hamiltonian for the electronic branches as
\begin{equation}
{\mathcal H}_e = \delta + u_0 + \hbar \omega_e(c^{\dagger}c + 1/2),
\end{equation}
where
\begin{equation}
\omega_e = \frac{2}{\hbar}\sqrt{\mu_1^2 - \mu_2^2},
\end{equation}
and $\beta=eB/\sqrt{m_x^em_y^e}$.
A similar approach yields, for the hole block,
\begin{equation}
{\mathcal H}_h = -\delta + u_0 - \hbar \omega_h(d^{\dagger}d + 1/2),
\end{equation}
where the $d$ operators are obtained from the $\alpha$ ladder operators via another Bogoliubov transformation, and
where
\begin{equation}
\omega_h = \frac{2}{\hbar}\sqrt{\lambda_1^2 - \lambda_2^2},
\end{equation}
and $\beta=eB/\sqrt{m_x^hm_y^h}$,
with
\begin{equation}
\lambda_1 = \eta_- -\frac{\chi^2}{4\delta}, \qquad \lambda_2 = \Delta_-+\frac{\chi^2}{4\delta}.
\end{equation}
The spectra obtained from Eq. (31) is shown as dashed red lines in Fig. 3 for Landau indices $n = 0,$ to $6$.
Similar expressions for the Landau levels in single layer phosphorene where obtained recently by means of a perturbative calculation in ref. \cite{Zhou}, which was based on the same tight-binding model employed here. However, in contrast with these results the present approach can be readily generalized for the bilayer case, as we show below.

\section{Bilayer phosphorene}

For the case of two coupled phosphorene layers, one now has to consider $8$ sublattices, which we label $A,B,C,D$ for the lower layer and $A',B',C'$ and $D'$ for the upper one. Using the tight-binding model of ref. \cite{Rudenko} one obtains the following Hamiltonian
\begin{equation}
{\mathcal H}_k=
\begin{pmatrix}
  H_1& H_c \\
  H_c & H_2
\end{pmatrix},
\end{equation}
acting on the spinors $\Psi = [\phi_A \quad \phi_B \quad \phi_D \quad \phi_C \quad \phi_A' \quad \phi_B' \quad \phi_D' \quad \phi_C']^T$,
where the $H_{1,2}$ blocks contain the interaction terms connecting sublattice sites within the same layer,
\begin{equation}
H_{1,2}=
\begin{pmatrix}
  u_{1,2} & t_{AB}(k)& t_{AD}(k) & t_{AC}(k) \\
  t_{AB}(k)^* & u_{1,2} & t_{AC}(k)^* & t_{AD}(k) \\
  t_{AD}(k) & t_{AC}(k) & u_{1,2} & t_{AB}(k)\\
  t_{AC}(k)^* & t_{AD}(k) & t_{AB}(k)^* & u_{1,2}
\end{pmatrix},
\end{equation}
with $u_{1,2}$ being the onsite energies for upper (1) and lower (2) layers. The $H_c$ blocks contain the couplings between sites located in adjacent layers; here, these correspond to the sublattice sites $A$, $B$, $C'$ and $D'$ with
$[{\mathcal H}_c]_{13}=t_{AD'}(k)$, $[{\mathcal H}_c]_{14}=t_{AC'}(k)$, $[{\mathcal H}_c]_{23}=t_{BD'}(k)=t_{AC'}(k)^*$ and $[{\mathcal H}_c]_{24}=t_{BC'}(k)=t_{AD'}(k)$, with the remaining elements being zero. The expressions for the coupling terms are given in the appendix.   
In the continuum approximation, the coupling terms become
\begin{eqnarray}
t_{AB}(k)&=&\delta_1 + \gamma_1k_x^2+ \gamma_2k_y^2+i\chi_1 k_y,\cr
t_{AC}(k)&=&\delta_2 + \gamma_3k_y^2+i\chi_2 k_y,\cr
t_{AD}(k)&=&\delta_3 + \eta_1k_x^2+ \eta_2k_y^2,\cr
t_{AD'}(k)&=&\delta_4 + \eta_3k_x^2+ \eta_4k_y^2,\cr
t_{AC'}(k)&=&\delta_5 + \gamma_4k_x^2+ \gamma_5k_y^2+i\chi_3 k_y.
\end{eqnarray}
where $\delta_1 =-2.85$ eV, $\delta_2 =3.61$ eV, $\delta_3 =-0.42$ eV, $\delta_4 =-0.06$ eV, $\delta_5 =0.41$ eV, $\eta_1 =0.58$ eV$\cdot$ \AA$^2$, $\eta_2 =1.01$ eV$\cdot$ \AA$^2$,
$\gamma_1 =3.91$ eV$\cdot$ \AA$^2$, $\gamma_2 =4.41$ eV$\cdot$ \AA$^2$, $\gamma_3 =-0.58$ eV$\cdot$ \AA$^2$, $\chi_1 =2.41$ eV$\cdot$ \AA,$\chi_2 =2.84$ eV$\cdot$ \AA, 
$\eta_3 =3.31$ eV$\cdot$ \AA$^2$, $\eta_4 =0.14$ eV$\cdot$ \AA$^2$,
$\gamma_4 =-0.56$ eV$\cdot$ \AA$^2$, $\gamma_5 =1.08$ eV$\cdot$ \AA$^2$, and 
$\chi_3 =1.09$ eV$\cdot$ \AA.

The above Hamiltonian leads to a system of $8$ coupled equations. However, as we show below, one can still obtain approximate analytical solutions. The eigenvalue equation can be rewritten as
\begin{equation}
\begin{pmatrix}
H_p & H_c' \\  
H_c' & H_m  
\end{pmatrix}
 = E\Psi' 
\end{equation}
where
\begin{equation}
H_p=
\begin{pmatrix}
H_{0}+H_{2}-\frac{1}{2}H_{3} & -i\frac{\Delta}{2}{\mathbf 1} \\  
 i\frac{\Delta}{2}{\mathbf 1} & H_{0}+H_{2}+\frac{1}{2}H_{3}
\end{pmatrix}
,
\end{equation}
\begin{equation}
H_m=
\begin{pmatrix}
H_{0}-H_{2}-\frac{1}{2}H_{3} & -i\frac{\Delta}{2}{\mathbf 1} \\  
 i\frac{\Delta}{2}{\mathbf 1} & H_{0}-H_{2}+\frac{1}{2}H_{3}
\end{pmatrix}
,
\end{equation}
and 
\begin{equation}
H_c'=
\begin{pmatrix}
-\frac{1}{2} H_{3} & 0 \\  
0 & \frac{1}{2}H_{3}
\end{pmatrix}
,
\end{equation}
where ${\mathbf 1}$ is the $2\times 2$ unit matrix, $\Delta$ denotes $u_1 - u_2$ and we assume $u_2=-u_1$, and
\begin{equation}
H_0=
\begin{pmatrix}
  0 & t_{AB}(k) \\
  t_{AB}(k)^* & 0
\end{pmatrix},
\end{equation}
\begin{equation}
H_2=
\begin{pmatrix}
  t_{AD}(k) & t_{AC}(k) \\
  t_{AC}(k)^* & t_{AD}(k)
\end{pmatrix},
\end{equation}
and
\begin{equation}
H_3=
\begin{pmatrix}
  t_{AD'}(k) & t_{AC'}(k) \\
  t_{AC'}(k)^* & t_{AD'}(k)
\end{pmatrix},
\end{equation}
and the eigenvectors are the $8$-component spinor
$\Psi'= [\psi_{pp} \quad \psi_{mp} \quad \psi_{pm} \quad \psi_{mm}]^T$ in which the four sets of $2$-component spinors are
\begin{eqnarray}
\psi_{pp}&=&
\frac{1}{2\sqrt{2}}
\begin{pmatrix}
  \phi_A+\phi_D+\phi_{A'}+\phi_{D'}\\  
  \phi_B+\phi_C+\phi_{B'}+\phi_{C'}
\end{pmatrix}
\cr
&&\cr
&&\cr
\psi_{pm}&=&
\frac{1}{2\sqrt{2}}
\begin{pmatrix}
  \phi_A-\phi_D-\phi_{A'}+\phi_{D'}\\  
  \phi_B-\phi_C-\phi_{B'}+\phi_{C'}
\end{pmatrix}
\cr
&&\cr
&&\cr
\psi_{mp}&=&
\frac{i}{2\sqrt{2}}
\begin{pmatrix}
  \phi_A+\phi_D-\phi_{A'}-\phi_{D'}\\  
  \phi_B+\phi_C-\phi_{B'}-\phi_{C'}
\end{pmatrix}
\cr
&&\cr
&&\cr
\psi_{pm}&=&
\frac{i}{2\sqrt{2}}
\begin{pmatrix}
  \phi_A-\phi_D+\phi_{A'}-\phi_{D'}\\  
  \phi_B-\phi_C+\phi_{B'}-\phi_{C'}
\end{pmatrix}
,
\end{eqnarray}
A further approximation can be made by taking into account the fact that the off-diagonal blocks $H_c'$
give rise to a small perturbation to the spectrum and can thus be neglected in a first approximation, leading to the following pair of eigenvalue equations:
\begin{equation}
\begin{pmatrix}
{H}_{0}+{ H}_{2}+\frac{1}{2}{H}_{3}-E & -i\frac{\Delta}{2}{\mathbf 1} \\  
  i\frac{\Delta}{2}{\mathbf 1} & {H}_{0}+{H}_{2}-\frac{1}{2}{H}_{3}-E
\end{pmatrix}
\begin{pmatrix}
  \psi_{pp}\\  
  \psi_{mp}
\end{pmatrix}
=0,
\end{equation}
and
\begin{equation}
\begin{pmatrix}
{H}_{0}-{H}_{2}+\frac{1}{2}{H}_{3}-E & -i\frac{\Delta}{2}{\mathbf 1} \\  
  i\frac{\Delta}{2}{\mathbf 1} & {H}_{0}-{H}_{2}-\frac{1}{2}{H}_{3}-E
\end{pmatrix}
\begin{pmatrix}
  \psi_{pm}\\  
  \psi_{mm}
\end{pmatrix}
=0.
\end{equation}
In this case, by solving Eq. (48) one obtains the $4$ inner families of branches (i.e. closer to the Fermi energy) whereas Eq. (49) leads to the outer families of levels. 

\begin{figure}
\includegraphics[width=7cm]{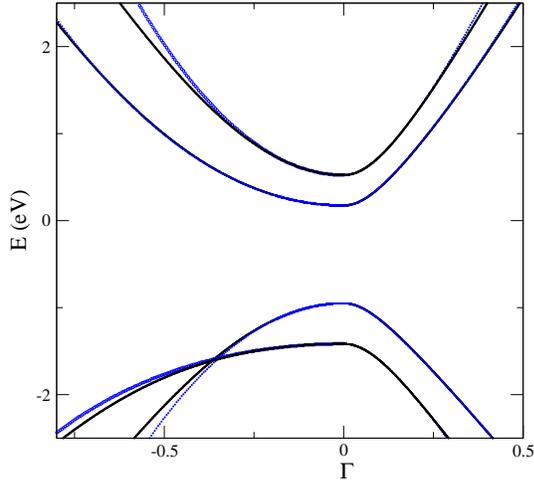}
\caption{\label{fig4} (Color online) Band structure of bilayer black phosphorus at the vicinity of the $\Gamma$ point, obtained form a tight-binding model (black solid lines) and the continuum approach (blue circles). }
\end{figure}

\begin{figure}
\includegraphics[width=7cm]{fig5.eps}
\caption{\label{fig5} (Color online) Low energy espectrum of bilayer phosphorene as function of the energy difference between layers obtained from Eq. (37) (black solid lines) and the reduced Hamiltonian Eq. (48) (blue dashed lines). }
\end{figure}

\begin{figure}
\includegraphics[width=7cm]{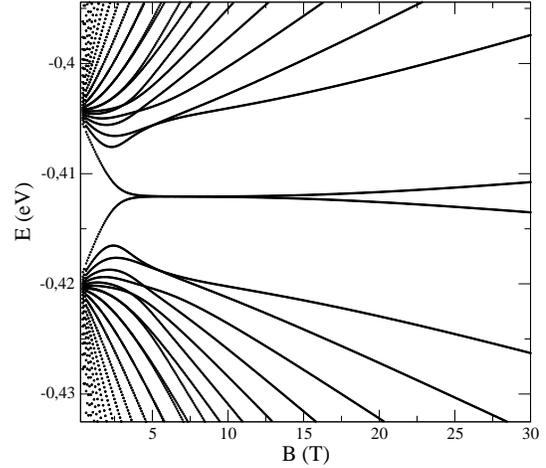}
\caption{\label{fig6} Landau levels as function of magnetic field, for $U_1 = 0.74$ eV and $U_2 = -U_1$. }
\end{figure}

\begin{figure}
\includegraphics[width=7cm]{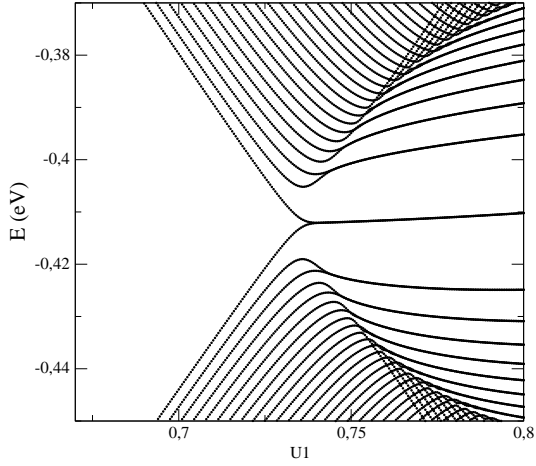}
\caption{\label{fig7} Landau levels as function of on-site energy, for $B = 10$ T, and $U_2 = -U_1$. }
\end{figure}

In the absence of bias, these systems of equations are reduced to $4$ copies of Eq. (9) although with different parameters. The resulting $8$ bands are labeled $i$,...$viii$ and the parameters corresponding to the four low-energy branches are shown in Table I, with the effective masses given as multiples of the electron mass $m_0$, with indices in decreasing order of energy.
For finite bias, the Hamiltonians Eq. (47) and (48) can be diagonalized. In order to show that, let us first recall Eq. (16) and rewrite the 2$\times$2 diagonal blocks in Eq. (47) as
\begin{equation}
H_0+H_2+\frac{1}{2}H_3=
\begin{pmatrix}
  \epsilon_1' & \epsilon_2' e^{i\theta_k'}  \\
\epsilon_2' e^{-i\theta_k'} & \epsilon_1'
\end{pmatrix}\quad , \quad
\end{equation}
and
\begin{equation}
H_0+H_2-\frac{1}{2}H_3=
\begin{pmatrix}
  \epsilon_1'' & \epsilon_2'' e^{i\theta_k''}  \\
\epsilon_2'' e^{-i\theta_k''} & \epsilon_1''
\end{pmatrix}\quad , \quad
\end{equation}
with the $\epsilon_{1,2}'$, $\epsilon_{1,2}''$ and the polar angles are defined as in Eqs. (17)-(19). Thus, after some straightforward algebra, we can obtain the four energy bands associated with Eq. (47) as the solutions of the equation
\begin{eqnarray}
&&[(E-\epsilon_1')^2-(\epsilon_2')^2][(E-\epsilon_1'')^2-(\epsilon_2'')^2]=-\Big(\frac{\Delta}{2}\Big)^4\cr
&&\cr
&&+\frac{\Delta^2}{2}\Big[\epsilon_2'\epsilon_2''\cos{(\theta_k'-\theta_k'')}+(E-\epsilon_1')(E-\epsilon_1'')\Big].
\end{eqnarray} 
For the range of energy and momenta of interest, one can safely assume $\cos{(\theta_k'-\theta_k'')} \approx 1$.  
In that case, Eq. (51) becomes
\begin{eqnarray}
&&[(E-\epsilon_1'-\epsilon_2')(E-\epsilon_1''-\epsilon_2'')-\Big(\frac{\Delta}{2}\Big)^2]\cr
&&\cr
&&\times [(E-\epsilon_1'+\epsilon_2')(E-\epsilon_1''+\epsilon_2'')-\Big(\frac{\Delta}{2}\Big)^2]=0.
\end{eqnarray} 
One can then obtain expressions for the energies of the low-energy bands at the $\Gamma$ point as function of $\Delta$ as
\begin{eqnarray}
E_c = \delta_1+\delta_2+\delta_3-\sqrt{\Big(\frac{\delta_4+\delta_5}{2}\Big)^2+\Big(\frac{\Delta}{2}\Big)^2}\cr
E_v = -\delta_1-\delta_2+\delta_3+\sqrt{\Big(\frac{\delta_4-\delta_5}{2}\Big)^2+\Big(\frac{\Delta}{2}\Big)^2}.
\end{eqnarray}

{\it Eigenstates:} Plane-wave eigenstates for the inner bands can be obtained from the Hamiltonian Eq. (47) as, for the conduction band:
\begin{equation}
\Psi_c(k)=A_c
\begin{pmatrix}
  1   \\
  a_ce^{-i\theta_k'} \\
  b_c \\
  c_ce^{-i\theta_k''}
\end{pmatrix}e^{i{\mathbf k}\cdot {\mathbf r}},
\end{equation}
with
\begin{equation}
a_c = \frac{(E-\epsilon_1')}{\epsilon_2'}+ i\frac{\Delta}{2\epsilon_2'}b_c,
\end{equation}
\begin{equation}
c_c = \frac{(E-\epsilon_1'')}{\epsilon_2''}b_c- \frac{\Delta}{2\epsilon_2'},
\end{equation}
and
\begin{equation}
b_c = \frac{2}{\Delta}\frac{[(E-\epsilon_1')^2-\epsilon_2'^2+\Delta^2 h/4]}{[E-\epsilon_1'+h(E-\epsilon_1'')]},
\end{equation}
with 
\begin{equation}
h = \frac{\epsilon_2'}{\epsilon_2''}e^{i(\theta_k'-\theta_k'')},
\end{equation}
and the other terms defined as before. It can be easily seen that, as $\Delta \rightarrow 0$ we obtain $a \rightarrow \pm 1$, $b,c \rightarrow 0$, as expected. For the valence band, the result is similar, with
\begin{equation}
\Psi_v(k)=A_v
\begin{pmatrix}
  b_v   \\
  c_ve^{-i\theta_k'} \\
  1 \\
  a_ve^{-i\theta_k''}
\end{pmatrix}e^{i{\mathbf k}\cdot {\mathbf r}},
\end{equation}
where
\begin{equation}
a_v = -\frac{(E-\epsilon_1'')}{\epsilon_2''}+ i\frac{\Delta}{2\epsilon_2''}b_v,
\end{equation}
\begin{equation}
c_v = \frac{(E-\epsilon_1'')}{\epsilon_2''}b_v- \frac{\Delta}{2\epsilon_2'},
\end{equation}
and
\begin{equation}
b_v = \frac{2}{\Delta}\frac{[(E-\epsilon_1'')^2-\epsilon_2''^2+\Delta^2 h'/4]}{[E-\epsilon_1''+h'(E-\epsilon_1')]},
\end{equation}
where $h'=1/h$. The normalizing constants are given by $A_{c,v}=[1+|a_{c,v}|^2+|b_{c,v}|^2+|c_{c,v}|^2]^{-1/2}$.
%%%%%%%%%

Figure 4 shows the spectrum of bilayer BP obtained from the tight-binding model (black solid lines) and continuum approaches (blue circles). As in the case of the single layer, the continuum results show a good agreement with the tight-binding data for the range $-1.5$ to $1.5$ eV. The effect of biasing on the gap is shown in Fig. 5 with data obtained from both the original $8\times 8$ tight-binding Hamiltonian (black solid lines) as well as from the analytical expression Eq.(54)
(blue dashed lines). The results show a good agreement, with a discrepancy of $\approx 4$ meV. For values of $\Delta$ above $\approx 1.5$ eV, the conduction and valence bands overlap, and the system becomes metallic.

{\it Landau levels}
The equations above lead to a set of $4$ electron and $4$ hole families of Landau level branches. In the absence of biasing (i.e. $\Delta = 0$), Eqs. (48) and (49) can be solved analytically in a similar fashion as in the case of single layer, with the parameters modified by the presence of interlayer coupling. Thus, the expressions for the different families of Landau level branches have the form
\begin{equation}
E = \delta_{\ell} \pm \hbar \omega_{\ell} (n+1/2),
\end{equation}
where
$\omega_{\ell}= eB/{\sqrt{m_x^{\ell}m_y^{\ell}}}$,
the ${\ell}$ indices denote different combinations of the coupling terms, with the positive sign corresponding to frequencies of electron branches (${\ell}=i,..., iv$)  and the negative sign is assigned to the hole branches (${\ell}=v,..., viii$). The values of $\delta_{\ell}$ and the effective masses are displayed in Table I. 

\begin{table}[!ht]
\caption{}
\begin{tabular}{|l|l|l|l|l|}
\hline
 $\ell$ & $iii$ & $iv$ & $v$ & $vi$ \\
\hline 
$\delta_{\ell}$ & $0.515$ eV    & $0.165$ eV   & $-0.947$ eV  & $-1.413$ eV  \\
\hline
$m_x^{\ell}$ & $0.65$ & $1.23$   & $0.72$   &  $2.74$ \\
\hline
$m_y^{\ell}$ & $0.17$  & $0.16$   &  $0.17$   & $0.18$ \\
\hline
\end{tabular}
\end{table}
In the presence of an external bias, a numerical approach becomes necessary. Figure 6 shows the dependence of the energy levels on the magnetic field, for a finite bias ($\Delta = 1.48$ eV). In this case, for the range $2$ T $\lessapprox B \lessapprox 12$ T the $n=0$ LL becomes doubly degenerate and weakly dependent on $B$, a situation that is analogous to the case of single layer graphene. This analogy is reinforced by the fact that the remaining levels become unevenly spaced.

The dependence of the Landau levels on the bias, for a fixed magnetic field, is shown in Fig. 7. It is seen that the presence of the external electric field tends to close the gap for a certain critical value of the bias. Moreover, the branches tend to become degenerate.

\section{Conclusions}

We have presented a continuum description of single layer and bilayer black phosphorus, starting from a tight binding model that reproduces the results of first principles calculations. Using this model we obtained the spectra of electrons and holes at the vicinity of the Fermi level at the gamma point and calculated the Landau level spectrum for both systems. For the case of bilayer BP we considered the effect of interlayer bias by introducing a layer-dependent on-site energy in the model. This showed that the presence of bias can close the electronic band gap, converting the material into a metal for a critical value of on-site energy difference. Correspondingly, the Landau level spectrum shows the appearance of doubly degenerate branches with a zero energy level weakly dependent of magnetic field for on-site energies above the critical value. This result agrees with recent {\it ab initio} calculations for few-layers black phosphorus \cite{Dolui} and can be exploited as the basis for future gate-tunable electronic devices. Furthermore, by taking into account additional interlayer hopping terms in a tight-binding description, the present model can be readily extended to deal with multilayer black phosphorus. 

\section{Acknowledgements}

J. M. Pereira Jr. acknowledges support from the Brazilian agency CAPES (Science Without Borders Program).

\section{Appendix}

The structure factors obtained from the tight-binding model of ref. \cite{Rudenko} are
given by the expressions
\begin{eqnarray}
t_{AB}(k) &=& 2t_1 \cos{(k_x a_1 \sin(\alpha_1 /2))}\times \cr
&&\exp[-ik_ya_1\cos{(\alpha_1 /2)}]\cr
&&{}\cr
&&+2t_3 \cos{(k_x a_1 \sin(\alpha_1 /2))}\times \cr
&&\exp[ik_y(a_1\cos{(\alpha_1 /2)}+2a_2\cos{\beta})]
\end{eqnarray}
\begin{eqnarray}
t_{AC}(k) &=& t_2 \exp[ik_ya_2\cos{\beta}]\cr
&&{}\cr
&&+t_5 \exp[-ik_y(2a_1\cos(\alpha_1 /2)+a_2\cos{\beta}]
\end{eqnarray}
\begin{eqnarray}
t_{AD}(k) &=& 4t_4 \cos{(k_x a_1 \sin(\alpha_1 /2))}\cr
&&\times\cos[k_y(a_1\cos(\alpha_1 /2)+a_2\cos(\beta)],
\end{eqnarray}
where $a_1$ is the distance between nearest neighbor sites in sublattices $A$ and $B$ or $C$ and $D$, and $a_2$ is the distance for n.n. sites of $A$ and $C$ or $B$ and $D$; $t_1$ and $t_2$ are the corresponding hopping parameters for nearest-neighbor couplings.
Due to the symmetry of the lattice, we have that $t'_{CD}(k) = (t'_{AB}(k))^*$, $t'_{CB}(k)=t_{AD}(k)$, $t'_{BD}(k)=(t'_{AC}(k))^*$, and $t_{BC}(k)=t_{AD}(k)$.
The bond angles are shown in Fig.1, and the parameters are $\alpha_1= 96,5^{\circ}$, $\alpha_2 = 101,9^{\circ}$, $\cos{\beta}=-\cos{\alpha_2}/\cos{\alpha_1}$ $a_1 = 2.22$ \AA, $a_2 = 2.24$ \AA. The hopping parameters are, in eV, $t_1 = -1.220$, $t_2 = 3.665$, $t_3 = -0.205$, $t_4 -0.105$, and $t_5. -0.055$  

For the case of bilayer BP, the additional coupling terms are
\begin{eqnarray}
t_{AD'}(k) &=& 4t_3^{\perp} \cos{(k_x 2a_1 \sin(\alpha_1 /2))}\times \cr
&&\cos{(k_y (a_1 \sin(\alpha_1 /2)+a_2\cos{\beta}))}\cr
&&+2t_2^{\perp}\cos{(k_y( a_1 \sin(\alpha_1 /2)+a_2\cos{\beta})},
\end{eqnarray}
and
\begin{eqnarray}
t_{AC'}(k) &=& 2t_1^{\perp} \cos{(k_x a_1 \sin(\alpha_1 /2))}\exp[ik_ya_2\cos{\beta}]\cr
&&{}\cr
&&+2t_4^{\perp}\cos{(k_x a_1 \sin(\alpha_1 /2))}\cr
&&\times\exp[-ik_y(2a_1 \sin(\alpha_1 /2)+a_2\cos{\beta})],
\end{eqnarray}
with $t_1^{\perp}=0.295$ eV, $t_2^{\perp}=0.273$ eV, $t_3^{\perp}=-0.151$ eV and $t_4^{\perp}=-0.091$ eV.

\end{document}